\documentclass[nonacm,manuscript,screen]{acmart}
\AtBeginDocument{%
  }

\setcopyright{none}
\begin{document}

\title{Interfacing Quantum Computing Systems with High-Performance Computing Systems: An overview}

%\email{webmaster@marysville-ohio.com}
%\affiliation{%
%  \institution{Institute for Clarity in Documentation}
%  \country{USA}
%}

\author{Konstantinos Rallis}
\orcid{0000-0003-0501-5554}
\affiliation{%
  \institution{National Centre for Scientific Research "Demokritos"}
  \city{Athens}
  \country{Greece}}
\affiliation{%
  \institution{Democritus University of Thrace}
  \city{Xanthi}
  \country{Greece}}
\email{krallis@ee.duth.gr}

\author{Ioannis Liliopoulos}
\orcid{0009-0006-2142-7670}
\affiliation{%
  \institution{National Centre for Scientific Research "Demokritos"}
  \city{Athens}
  \country{Greece}}
\affiliation{%
  \institution{Democritus University of Thrace}
  \city{Xanthi}
  \country{Greece}}
\email{ililiopo@ee.duth.gr}

\author{Georgios D. Varsamis}
\orcid{0000-0003-0464-4842}
\affiliation{%
  \institution{National Centre for Scientific Research "Demokritos"}
  \city{Athens}
  \country{Greece}}
\affiliation{%
 \institution{Democritus University of Thrace}
 \city{Xanthi}
 \country{Greece}}
 \email{gevarsam@ee.duth.gr}

\author{Evangelos Tsipas}
\orcid{0000-0002-4140-1308}
\affiliation{%
  \institution{National Centre for Scientific Research "Demokritos"}
  \city{Athens}
  \country{Greece}}
\affiliation{%
  \institution{Democritus University of Thrace}
  \city{Xanthi}
  \country{Greece}}
  \email{etsipas@ee.duth.gr}

\author{Ioannis G. Karafyllidis}
\orcid{0000-0003-2079-5480}
\affiliation{%
  \institution{Democritus University of Thrace}
  \city{Xanthi}
  \country{Greece}}
\email{ykar@ee.duth.gr}

\author{Georgios Ch. Sirakoulis}
\orcid{0000-0001-8240-484X}
\affiliation{%
  \institution{Democritus University of Thrace}
  \city{Xanthi}
  \country{Greece}}
\email{gsirak@ee.duth.gr}

\author{Panagiotis Dimitrakis}
\orcid{0000-0002-4941-0487}
\affiliation{%
  \institution{National Centre for Scientific Research "Demokritos"}
  \city{Athens}
  \country{Greece}}
\email{p.dimitrakis@qi.demokritos.gr}

\renewcommand{\shortauthors}{Rallis et al.}

\begin{abstract}
  The connection and eventual integration of High-Performance Computing (HPC) with Quantum Computing (QC) represents a transformative advancement in computational technology, promising significant enhancements in solving complex, previously intractable problems. This manuscript provides a comprehensive overview of the current state of HPC-QC interfacing, detailing architectural methodologies, software stack developments, middleware functionalities, and hardware integration strategies. It critically assesses existing hardware-level integration models, ranging from standalone and loosely-coupled architectures to tightly-integrated and on-node systems. The software ecosystem is analyzed, highlighting prominent frameworks such as Qiskit, PennyLane, CUDA-Q, and middleware solutions like Pilot-Quantum, essential for seamless hybrid computing environments. Furthermore, the manuscript discusses practical applications in optimization, machine learning, and many-body dynamics, where hybrid HPC-QC systems can offer substantial advantages. It also describes existing challenges, including hardware limitations (coherence, scalability, connectivity), software maturity, communication overhead, resource management complexities, and cost factors. Finally, future directions towards tighter hardware and software integration are discussed, emphasizing ongoing research developments and emerging trends that promise to expand the capabilities and accessibility of hybrid HPC-QC systems.
\end{abstract}



\maketitle
\section{Introduction}
\label{sec:intro}

As computing technology continues to evolve, it is increasingly applied to a broader range of problems in both research and industry, as well as in everyday life. This technological expansion is inevitably accompanied by a growing demand for computational resources to address increasingly complex and data-intensive large-scale problems. To cover the growing demands for computational resources and computation power, High-Performance Computing (HPC) systems have been employed, and so far, they are the dominant approach for dealing with large-scale computations with enhanced performance. Such systems are now extensively used not only by research centers and organizations, but also by industry sectors. Enterprises, governmental bodies, and international collaborations have driven the development, deployment, and enhancement of HPC infrastructures by funding and operating their own systems. At the same time, significant research efforts have focused on optimizing performance, improving energy efficiency, enhancing scalability, and increasing accessibility of these systems.

In parallel, apart from the progress and advancements in High-Performance Computing, novel computing systems have emerged and have already attracted significant research and implementation efforts: Quantum Computers. This technology promises to provide a significant boost to the available computational power and to significantly enhance the problem solving capabilities of the existing systems. By exploiting the fundamental quantum mechanical phenomena that govern the operation of quantum computers, these systems are expected to tackle computationally intensive problems, such as a set of specific NP-hard and NP-complete problems, more efficiently \cite{au2023np}.

However, due to current hardware limitations, the NISQ (Noisy Intermediate-Scale Quantum) era quantum computers cannot yet be used for large-scale problems. 
They are currently still requiring specific installation standards and infrastructure with demanding maintenance. As of today, several big tech companies provide their own quantum computers for use, and with the evolution of the quantum hardware technology, more and more quantum resources with increased computational capabilities (increased number of qubits, the unit of quantum computation) are becoming available to the public \cite{alvarado2024technological}.

Running a program and solving a problem on a quantum computer while also harvesting its special capabilities and achieving actual performance gains is not completely straightforward. It requires special handling and encoding of data, as well as an appropriate formulation of the problem in order to be solved by a quantum computer \cite{zhu2025quantum}. Thus far, there is no single standardized compilation process that automatically formulates and maps every problem to a QC-compatible form, making this an active area of ongoing research.

As previously mentioned and evident from the above, today’s NISQ-era quantum computers cannot tackle large-scale problems and require an additional classical system to handle essential computations. As of today, the most commonly accepted way of using a QC system is to have it interfaced and in combination with a classical computing system, mainly an HPC system. In this approach, quantum computers and their computing part, the Quantum Processing Unit (QPU), are treated by the HPC system as another computing unit (resource), just like CPU and GPU nodes in heterogeneous distributed systems \cite{elsharkawy2023integration}. Such an approach, even though promising, introduces a set of hard-to-solve challenges. The HPC system is responsible for formulating, encoding and mapping the problem for the QPU, assign jobs to QPUs and time-manage their operation while also preparing and fetching the required data and instructions to them. This integration of QPUs to HPC computing systems and the effective HPC-QC interfacing are crucial steps for the exploitation of current QC systems, their use in actual problems where they can provide actual advantages in terms of performance, as well as for the development of the QC technology as a whole.

In this work, we present a comprehensive review of state-of-the-art methodologies and emerging trends in the integration of High-Performance Computing (HPC) with Quantum Computing (QC), structured as follows: In Section~\ref{sec:background}, we provide essential background on HPC and QC technologies, as well as on hybrid HPC-QC systems, highlighting their characteristics and capabilities. Section~\ref{sec:softwaremiddleware} delves into the software ecosystem, detailing key quantum programming frameworks and middleware solutions, including main representatives such as Qiskit, PennyLane, CUDA-Q, Pilot-Quantum and others. In Section~\ref{sec:hardware_interface}, we discuss hardware-level integration strategies, covering physical interconnect technologies and quantum networking. Section~\ref{sec:applications} illustrates practical application domains where hybrid HPC-QC systems have shown significant promise, specifically optimization problems, quantum machine learning, and quantum simulations. Then in Section~\ref{sec:challenges}, we critically examine existing technical and operational challenges that currently limit broader adoption, and finally summarize and conclude in Section~\ref{sec:conclusions}, commenting on possible future research directions.

%%%%%%%%%%%%%%%%%%%%%%%%%%%%%%%%%%%%%%%%%%%%%%%%%%%%%%%%%%%%%%%%%%%%%%%%%%%%%%%%%%%%%%%%%%%%%%%%%%%%%%%%%%%%%%%%%%%%%%%%%%

\section{Background}
\label{sec:background}
\subsection{High-Performance Computing Systems} 
\label{subsec:hpcs}
High-Performance Computing (HPC), is a technology that goes far beyond the classical personal computing systems as we know them, not only in terms of their form factor, but also in terms of their operation use. At their core, HPC systems are based on the same operational principles as conventional computing systems. They rely on digital logic circuits and execute Boolean algebra, whose basic computational unit is the binary digit bit, and their basic computing cell unit is the transistor. However, HPC is more than that. It is a technology that uses a set of computing resources, clusters of processing units. Those processing units can be up to several thousands in a single HPC system, and they can also be of a heterogeneous nature. This means they do not only consist of CPUs, but they can also contain other types of specialized accelerators, such as GPUs, especially suitable for parallel processing workloads, FPGAs, TPUs, Neuromorphic Processors and other units, which can be used to handle several diverse tasks \cite{milojicic2021future, macia2024towards, rajan2024design}.

Such systems are mainly developed by big tech companies and even countries that are able to handle the costs of manufacturing and maintenance. In order to cover the increasing needs of computing resources in an affordable manner, HPC resources are also provided to the public through the cloud, in the form of service \cite{hoefler2024xaas}.

Nowadays, in the era of Artificial Intelligence (AI), HPC systems are becoming more and more significant and required by a greater pool of users of diverse backgrounds, as they find application in a broader set of fields with increased requirements in computational power and resources. They are the driving force of research in AI, as they can handle through their resources the computing-intensive tasks of training of large scale AI models and preprocessing and manipulation of Big Data \cite{yin2022comparative, ettifouri2024need}. HPC systems are also highly used in the Automotive Industry, where intensive simulation is required for the design and parameter estimation of new products, among others Computational Fluid Dynamic (CFD) simulations for Aerodynamics \cite{galeazzo2024performance}. Related to that, HPC systems are also used in the field of Energy prediction and storage, including weather forecasting, especially nowadays where renewable energy sources are dominant and are heavily based on external environmental factors (solar irradiance, wind etc.) \cite{nakaegawa2022high}. They are an indispensable tool in the hands of governments as means for civil protection, used to predict possible natural disasters (i.e. seismic activity, tidal waves, tsunamis etc.) \cite{gribova2024information}. HPC systems can also be used for tasks related to banking and financing \cite{liu2024fingpt, song2022design} or even, through their heterogeneous resources, in the form of render farms for 3D Visuals \cite{kolarek2024cultural}. All this is just a representative subset of the full application spectrum of HPC systems.

\subsection{Quantum Computing Systems}
\label{susec:qcs}
Quantum computing and quantum computers represent a different approach to computing compared to what has been reported above. A quantum computer is practically a machine that performs computations by leveraging quantum mechanical phenomena, such as superposition and quantum entanglement. The basic unit of quantum computing is the qubit (quantum bit), which, unlike classical bits that can represent information as binary values, is capable of encapsulating not only two values, $0$ and $1$, but also infinite values in between, in a form of coherent superposition between those two "basis" states. Moreover, when two or more qubits become entangled, their quantum states become related and interdependent, enabling complex quantum operations that can lead to actual computational advantage \cite{rietsche2022quantum}.

In terms of representing those states and thus physically fabricating the qubits in hardware, there is no single standardized approach dominating the field yet. Several alternative quantum computing hardware technologies have already been proposed and are in active development and investigation. The most common of them include superconducting qubits \cite{bravyi2022future}, trapped-ion qubits \cite{brown2021materials}, photonic (light-based) qubits \cite{kok2007linear}, neutral atoms \cite{cohen2021quantum}, and spin qubits in silicon \cite{burkard2023semiconductor}. Several technology companies and research groups around the world are investigating and investing in one or more of those technologies, each one providing its own characteristics and having its own advantages and disadvantages as shown in Tab.~\ref{tab:qubit_comparison}, targeting to achieve scalable and fault-tolerant quantum computation.

Quantum computing promises to substantially accelerate solutions to specific classes of problems that even classical HPCs would not be able to solve in feasible time. In fact, what quantum computers are able to do, is to solve specific types of computationally intensive problems, many of which are lying in the NP-complete and NP-hard categories, more efficiently, in some cases even in polynomial time \cite{au2023np}. Such problems, on which quantum computing can be beneficial, are present in a wide variety of fields. Healthcare and more specifically drug discovery is one of the most common, referring mainly to the speedup of protein folding simulations and the investigation of drug target interactions \cite{wang2023recent,varsamis2023quantum,liliopoulos2025quantum}. Beyond protein folding, quantum algorithms find application in genomics-based diagnostics, personalized medicine and pathogen detection, through tasks such as DNA sequence alignment and genome assembly \cite{varsamis2023quantumalgo,varsamis2023quantumgate}. Adding to that, QC can significantly aid Climate Simulations, as climate and weather are highly complex many-dimensional systems and simultaneously help to the highly related task of Energy Grid optimization \cite{blenninger2024q}. Another field of application for QC is Cryptography and Security, mainly based on its capabilities to deliver fast solutions to the factorization problem through the well-known Shor's algorithm \cite{ugwuishiwu2020overview}. QCs are practically creating a new whole segment in Cryptography, called Post-Quantum Cryptography. Machine Learning and AI are additional areas where the application of QCs is tested, mainly through Quantum Machine Learning, especially for tasks with large feature spaces. QCs can also be exploited for their ability to simulate quantum systems (Quantum Simulation), mainly targeting Material Science, Chemistry and many more \cite{daley2022practical}. It also offers significant boost to combinatorial problems through enhanced combinatorial optimization \cite{chicano2025combinatorial} and excels in searches in unsorted databases mainly through Grover's Algorithm \cite{qiu2024distributed}.

But besides its prospects and seemingly wide spectrum of application, QC is not a panacea. The speedup that they offer tends to be problem-specific, fitting specific classes of problems that can harness their special properties, like those that have been mentioned above \cite{gill2022quantum}. Being at a technology readiness level (TRL) of 3-4, and still in the NISQ era, QC technology has yet several significant challenges to overcome, such as error correction, qubit coherence, scalability and many more. Overcoming these obstacles is essential for quantum computing to transition from experimental demonstrations to practical, large-scale, and fault-tolerant computational platforms.

%%%%%%%%%%%%%%%%%%%%%%%%%%%%%%%%%%%%%%%%%%%%%%%%%%%%%%%%%%%%%%%%%%%%%%%%%%%%%%%%%%%%%%%%%%%%%%%%%%%%%%%%%

\begin{table*}[htbp]
    \caption{Comparison of different qubit technologies.}
    \label{tab:qubit_comparison}
    \centering
    %\renewcommand{\arraystretch}{1.1}
    %\begin{adjustbox}{width=\textwidth}
    {\small
    \begin{tabular}{|p{1.5cm}|p{2.1cm}|p{1.8cm}|p{1.8cm}|p{1.8cm}|p{1.8cm}|p{1.8cm}|}
    \hline
    \textbf{Category} & \textbf{Superconducting Qubits} & \textbf{Trapped-Ion Qubits} & \textbf{Photonic Qubits} & \textbf{Neutral-atom Qubits} & \textbf{Spin Qubits in Silicon} & \textbf{Topological Qubits} \\ \hline
    
    \textbf{Operation} &
    Zero-resistance superconductors at millikelvin temperatures (Josephson junctions). &
    Ions trapped by electromagnetic fields; quantum states encoded in energy levels. &
    Quantum information carried by single photons, manipulated/detected optically. &
    Neutral atoms in optical tweezers/lattices; Rydberg excitation for strong coupling. &
    Single-electron spins in silicon quantum dots; quantum control similar to gate-controlled transistors. &
    Quasi-particles (e.g., Majorana fermions). \\ \hline

    \textbf{Pros} &
    Mature fabrication, relatively fast gates. &
    Long coherence times. High-fidelity gate operations. All qubits are identical by nature. &
    Fast transmission, less prone to decoherence (weak interaction with the environment). &
    Highly scalable arrays, long coherence. &
    Uses mature silicon tech, potential for classical/quantum integration. &
    Potential intrinsic fault tolerance, reduced overhead for error correction. \\ \hline

    \textbf{Challenges} &
    Requires very low temperatures (deep cryogenics). &
    Slower gates, difficult scaling and multi-zone control. &
    Hard to make two-qubit gates, efficient and scalable on-chip photon sources and detectors, and optical loss, are still challenging. &
    Precise large-array control, high-gate fidelity stability and uniformity. &
    Demands extreme spin control, decoherence from nuclear spins. &
    Highly experimental, complex materials, fabrication and scalability unclear. \\ \hline

    \textbf{Key Players} &
    IBM, Google, Rigetti, academic/industrial consortia. &
    IonQ, Quantinuum (Honeywell), Alpine Quantum Technologies (AQT), various universities. &
    Xanadu, PsiQuantum, photonics research groups. &
    QuEra, Pasqal, Infleqtion (ColdQuanta). &
    Intel, Quantum Motion, UNSW (Australia). &
    Microsoft. \\ \hline
    \end{tabular}%
    }
    %\end{adjustbox}
\end{table*}

%%%%%%%%%%%%%%%%%%%%%%%%%%%%%%%%%%%%%%%%%%%%%%%%%%%%%%%%%%%%%%%%%%%%%%%%%%%%%%%%%%%%

\subsection{Interfacing quantum computers with high-performance computers}

High-performance computing (HPC) and quantum computing (QC) as presented previously, are two distinct yet complementary technological domains. Interfacing them into quantum-classical Hybrid Systems and leveraging their enhanced combined capabilities provides a great potential to revolutionize various fields, spanning from scientific research to complex industrial applications \cite{humble2016software}.

Quantum computers can perform certain computations exponentially faster than classical computers, like combinatorial optimization and factorization, unstructured search, and others, making them particularly well-suited for solving complex problems that are intractable for traditional high-performance computing systems. Integrating quantum computers with high-performance computing environments allows for a powerful hybrid approach, where classical systems are used to manage the overall workflow and offload to the QPUs specific tasks that are better suited for quantum processing \cite{chehimi2022physics}. At an abstract level, such a workflow will include initially the breaking down -by the HPC- of a complex problem to individual subtasks. The classical system will determine which one of those suits better to be executed by the HPC and which one can benefit from execution on a Quantum Processor. Then the classical system will again proceed with the whole orchestration process, the job scheduling and assignment, as well as the data management. This involves the encoding of data and mapping of the selected tasks to a form that a Quantum Processor is able to understand and process. After the execution of the each subtask by both the classical and quantum nodes, the classical system is again responsible for the handling, decoding and combination of the output data.

As quantum computers and thus quantum processors continue to evolve by increasing the number of fault tolerant qubits, the need to interface them with high-performance computing infrastructures has become increasingly important and promising for the development of the technology \cite{moller2017impact, britt2017high, soeparno2021cloud}. As mentioned in \cite{humble2016software}, a promising approach to address the challenges that arise with the design of large-scale quantum processors is to mimic modern high-performance computing systems, where thousands of heterogeneous processors in various forms and storage units are interconnected via a communication network, and the computational problems are solved by adopting a distributed computing approach \cite{humble2016software}. This will allow quantum processors to be seamlessly integrated into a high-performance computing infrastructure, in the form of additional distributed computing resources, similar to conventional processors, enabling the efficient and optimized distribution and parallelization of workloads between classical and quantum components.

Another key aspect of interfacing quantum computers with high-performance computing systems, is the development of sophisticated software tools and frameworks that can facilitate the integration and smooth operation of these two distinct computing paradigms. This step is crucial for valid mapping of computing problems in a form suitable for Quantum Computers, as well as for appropriate job scheduling and resource allocation, in order to avoid significant overheads and delays that will ultimately reduce the total performance of the Hybrid System. 

%%%%%%
Recent research has particularly focused on developing programming models and execution frameworks that abstract the complexity of quantum hardware, providing unified interfaces for developers to design and deploy hybrid quantum-classical applications. As described in \cite{elsharkawy2023integration}, integration can occur at multiple layers: at the application layer, classical and quantum algorithmic modules are combined via hybrid programming languages or workflow systems; at the compilation layer, hybrid source code is compiled into appropriate instructions; and at the hardware layer, HPC binaries and QPU pulse-level instructions are scheduled for execution on their target devices. Similarly, \cite{schulz2022accelerating} presents an approach for integrating HPC and QC software environments into a unified workflow, emphasizing the importance of bridging existing software stacks for seamless operation. Architectural solutions like the shared buffer memory space in \cite{mccaskey2018language} facilitate direct interaction between host CPUs and quantum processors, managed by accelerator systems and shared memory buffers. Other approaches, such as the unified HPC-QC toolchain discussed in \cite{seitz2023toward}, allow for high-level algorithm compilation, resource scheduling, and runtime orchestration across hybrid computational nodes. Finally, cloud-based integration models, reported by \cite{ravi2021quantum}, showcase the paradigm where clients submit quantum jobs to vendor-managed clouds, with jobs managed, queued, and executed remotely. These architectural approaches illustrate the diverse methods being explored to enable efficient and practical HPC-QC integration.
%%%%%%%%%

Several software-based approaches and frameworks have already been developed specifically to address the challenges of integrating quantum computing systems with high-performance computing infrastructures. Those software stacks cover the whole spectrum of topics that are required for an efficient co-existence of HPCs and QCs in Hybrid Systems, from programming and problem mapping to transpiling, scheduling, resource allocating and managing of input and output data. The aforementioned existing architectures for hybrid quantum-classical systems as well as the existing techniques and the software stack for HPC-QC interconnection, will be further analyzed and compared.

%%%%%%%%%%%%%%%%%%%%%%%%%%%%%%%%%%%%%%%%%%%%%%%%%%%%%%%%%%%%%%%%%%%%%%%%%%%%%%%%%%%%%%%%%%%%%%%%%

\section{Software Stack and Middleware}
\label{sec:softwaremiddleware}
\subsection{Programming Models}

Over the last few years there has been an extensive effort in order to design better and less noise-prone QCs and more efficient HPC systems \cite{reed2022reinventing,proctor2025benchmarking, hockings2025scalable, gustafson2022noise}. Furthermore, there is a drift among the scientific community towards the interconnection of those two systems via a low-latency physical link \cite{ruefenacht2022bringing}. Although this research field has drawn the attention of the majority of the scientific community, emphasis shall also be given in designing the appropriate software stack that enables anyone to exploit the computational power originating from the seamless integration of the aforementioned components \cite{schulz2022accelerating}.   

To achieve seamless integration between HPCs and QCs, a viable software stack must be able to combine and employ the benefits from two types of systems which are entirely different from one another. On one side, Quantum Computers exploit quantum phenomena like the Quantum Superposition and Quantum Entanglement in order to tackle hard-to-solve computational problems. Current state-of-the-art quantum computing frameworks provide a quite friendly end-user environment and allow users to program Quantum computers, typically in the form of constructing quantum circuits, by using high-level programming languages like Python or C++. These quantum circuits are then transpiled to be made compatible with the topology of a specific quantum device and optimized for execution on present day noisy quantum systems, a process known as routing \cite{cowtan2019qubit}. Of course, these tasks are almost impossible to be manually performed by users and thus all commercially available frameworks provide automatic tools to handle them. The most popular commercially available QC programming frameworks are summarized in Tab.~\ref{tab:qc_frameworks}.

\begin{table}[t!]
\caption{Most popular commercially available QC programming frameworks.}
\label{tab:qc_frameworks}
\centering
\resizebox{\textwidth}{!}{%
\begin{tabular}{|l|l|l|l|}
\hline
\textbf{Name} & \textbf{Vendor} & \textbf{Programming Language} & \textbf{Target Qubit Technology} \\
\hline
Qiskit \cite{javadi2024quantum} & IBM & Python & Superconducting \\
\hline
Classiq \cite{goldfriend2024design} & Classiq Technologies & Python/Qmod \cite{goldfriend2024design} & Superconducting, Trapped-Ion, Spin Qubits, Photonic \\
\hline
Cirq \cite{cirq_developers_2025_15191735} & Google & Python & Trapped-Ion, Neutral-Atom \\
\hline
PennyLane \cite{bergholm2018pennylane} & Xanadu & Python & Superconducting, Trapped-Ion, Neutral-Atom, Photonic etc. \\
\hline
CUDA-Q \cite{the_cuda_q_development_team_2025_15407754} & NVIDIA & C++/Python & Superconducting, Trapped-Ion, Neutral-Atom, Photonic \\
\hline
Braket \cite{awsBraket} & Amazon & Python/Julia & Superconducting, Trapped-Ion, Neutral-Atom \\
\hline
Azure Quantum \cite{Microsoft_Azure_Quantum_Development} & Microsoft & Python/Q\# \cite{svore2018q} & Superconducting, Trapped-Ion, Neutral-Atom, Topological (experimental) \\
\hline
Forest \cite{smith2016practical} & Rigetti & Python/Quil \cite{smith2016practical} & Superconducting \\
\hline
Qadence \cite{seitz2025qadence} & Pasqal & Python & Neutral-Atom \\
\hline
qBraid SDK \cite{hill_2024_12627597} & qBraid & Python & Superconducting, Trapped-Ion, Neutral-Atom \\
\hline
\end{tabular}
} % end resizebox
\end{table}

On the other side, HPCs comprise a vast number of computational nodes like CPUs, GPUs and FPGAs, which are working in parallel, in order to address many complex scientific challenges, like large-scale simulations and big data analytics etc. \cite{beck2024integrating}. Similarly to QCs, all programs sent for execution to HPCs must be precompiled. Furthermore, when programming large HPCs the concepts of job scheduling, node communication and synchronization must be taken into consideration. Job scheduling refers to the process of resource allocation and job distribution among the available computational nodes, aiming for optimal system performance. Additionally, due to the distributed “nature” of the HPCs, information must be exchanged among system’s nodes and all nodes must be perfectly synchronized, under the auspices of a central node, to ensure that all data dependencies are satisfied. The computational overhead added to the total execution time, due to the aforementioned processes, is proportional to the volume of the data shared and the inter-node communication frequency. Current HPC programming frameworks contain components for code compilation and runtime execution monitoring along with a broad range of libraries and interfaces for efficient application development and parallel programming, like OpenMP \cite{openmp2024api} and MPI \cite{mpi2023standard}. The most dominant high-level languages for HPC programming are C, C++ and Fortran, since the source code written in them is compiled before its execution.

\begin{figure}
    \centering
    \includegraphics[width=0.65\linewidth]{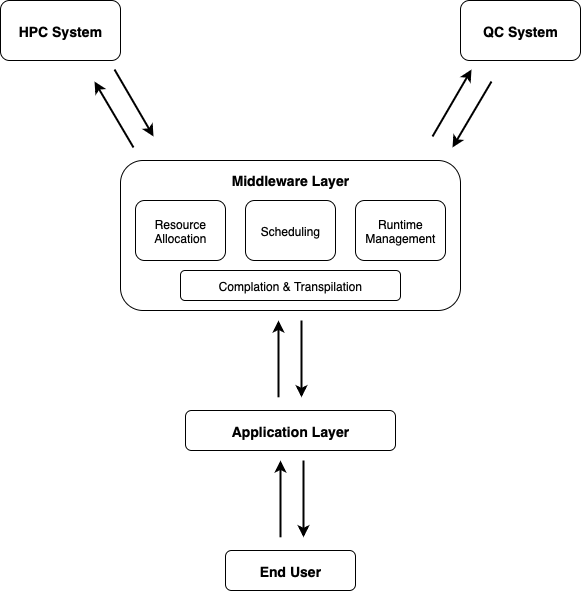}
    \caption{High-level architecture of a hybrid HPC-QC computing workflow. The application layer interacts with the end user and communicates tasks to the middleware layer, which is responsible for resource allocation, scheduling, compilation, and runtime management across both the high-performance computing (HPC) system and the quantum computing (QC) system. The middleware coordinates the seamless integration of classical and quantum resources to execute hybrid applications.}
    \label{fig:hpcqcworkflow}
    \Description{Diagram of a high-level hybrid HPC–quantum computing workflow. 
The end user interacts with the application layer, which sends tasks to the middleware layer. 
The middleware contains modules for resource allocation, scheduling, runtime management, and compilation and transpilation. 
It connects bidirectionally with both the high-performance computing system and the quantum computing system to coordinate execution of hybrid workloads.}

\end{figure}

Understandably, any software stack designed to link HPCs and QCs has to comprise all the aforementioned components to ensure a seamless integration between the two systems. A representative figure that illustrates the workflow of such a software stack is presented in Fig.~\ref{fig:hpcqcworkflow}.

Current state-of-the-art software stacks imitate one another regarding their abstraction layers. Most of them comprise three discrete layers, namely the ‘application’, ‘compilation’ and ‘hardware’ layer \cite{elsharkawy2024integration}. The latter two are usually merged in a single layer under the name ‘Middleware’ or ‘Orchestration’ layer. Its components and role will be analyzed in the following section. The ‘application’ layer usually supports a user-friendly front-end ecosystem that contains all the appropriate tools and functions for efficient code design (compilers, libraries etc.) \cite{meller2025programming}. In most cases this front-end is Python-driven and it comes with the appropriate bindings to other high-level languages (e.g. C/C++ or FORTRAN). At this level of abstraction, the software must support a plethora of QC programming frameworks \cite{humble2021quantum} and thus must be QC programming-agnostic.

\begin{figure*}[t!]
    \centering
    \includegraphics[width=1\textwidth]{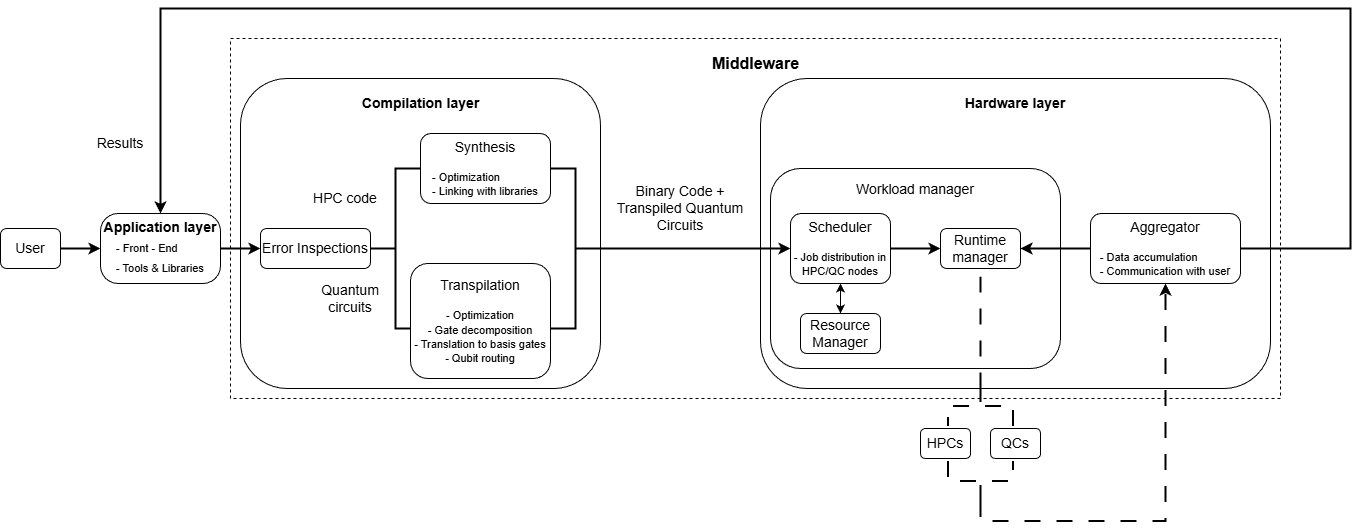}
    \caption{A higher-level abstraction of the ‘Middleware’ architecture.}
    \label{fig:middlewarearch}
    \Description{Block diagram of a middleware architecture for HPC–quantum integration. 
The workflow starts with the application layer, where the user accesses front-end tools and libraries. 
Output passes through an error inspection step and into the compilation layer, which includes synthesis for optimization and library linking, and transpilation for gate decomposition, translation to basis gates, and qubit routing. 
The resulting binary code and transpiled quantum circuits are sent to the hardware layer’s workload manager, which contains a scheduler for distributing jobs to HPC and quantum computing nodes, a resource manager, a runtime manager, and an aggregator that accumulates data and communicates with the user.}

\end{figure*}

\subsection{Middleware/Orchestration Frameworks}

The ‘Middleware’ layer or just ‘Middleware’ is the most important component in an HPC-QC software stack since it is responsible for the compilation of the hybrid workload, the proper job scheduling, the monitoring of the heterogenous (classical and quantum) tasks and the information interchange between the HPC nodes and the QCs. As mentioned above, the ‘Middleware’ consists of two main layers, the ‘compilation’ and the ‘hardware’ layer. From a higher-level abstraction the ‘Middleware’ architecture is demonstrated in Fig.~\ref{fig:middlewarearch}.  

When in the ‘compilation’ layer the hybrid code undergoes a series of error inspections (lexical, semantic etc.) and then is separated in its classical and quantum counterpart. Then, the classical (i.e. the source code to be executed in HPCs) code is optimized, during the synthesis stage, and combined alongside the appropriate library files to form the binary file. The quantum circuits (i.e. the quantum part) are also optimized and transpiled in terms of the target QC. After that, the binary file(s) and the transpiled quantum circuit(s) are forwarded to the ‘hardware’ layer to be executed in the corresponding nodes. 

The ‘hardware’ layer is in charge of the workload execution lifecycle. It consists of two modules, namely the ‘workload manager’ and the ‘aggregator’. The ‘workload manager’, itself, consists of three submodules, namely the ‘scheduler’, ‘resource manager’ and ‘runtime manager’. Initially, the combined workload from the ‘compilation’ layer is transferred to the ‘scheduler’, which distributes it to the appropriate HPC and QC nodes for execution, via the ‘runtime manager’. To achieve that, schedulers are in constant communication with ‘resource managers’, which are aware for the current status of each node. The process of job scheduling is quite trivial, as many parameters must be taken into account, such as each node’s idle time, each process’ execution time and possible data dependencies between jobs that may lead to further communication overhead \cite{fan2021job}, to name a few. The ‘runtime manager’, as the name suggests, is responsible for the program execution on both classical and quantum hardware and for the inter-node communication and synchronization. Some of the most commonly-used workload managers are SLURM \cite{yoo2003slurm} and Agnostiq’s Covalent \cite{will_cunningham_2025_15400489}. Last but not least, the ‘aggregator’ module collects the execution results from both classical and quantum hardware. Then it forwards them to both the ‘runtime manager’ so that any inter-node data dependencies to be satisfied and all nodes to be synchronized and to the application layer for demonstration purposes.

\subsection{Existing Platforms and Frameworks} 

In this section, we will discuss and refer to both existing QC frameworks that support the appropriate tools for HPC integration and state-of-the-art HPC-QC platforms and frameworks designed in a software-agnostic manner to support multiple QC hardware. The most well-known QC frameworks that support HPC integration tools are the following. 

\textbf{Qiskit \cite{javadi2024quantum}:} Qiskit is a Python-based open-source framework, designed by IBM. It is one of the most-favorable QC programming and algorithm design frameworks, due to its simplicity and vast number of tools available. It comprises many simulators, both ideal and with noise, which enable users to test their quantum algorithms behavior under NISQ devices. Furthermore, it includes tools which allow users to execute their quantum circuits on real IBM’s quantum computers and optionally to improve their results by applying error mitigation and suppression techniques \cite{ezzell2023dynamical,van2022model,wallman2016noise}. Additionally, it provides some interfaces, like the Qiskit Serverless, for continuous HPC-QC integration and resource management, through cloud-based services in the IBM Quantum Platform. As of April 2025, Qiskit (v2.x) also supports limited C-based QC programming. 

\textbf{PennyLane \cite{bergholm2018pennylane}:} PennyLane is a Python-based open-source framework, designed by Xanadu. Like Qiskit, it is one of the most used QC programming frameworks, especially in terms of Quantum Machine Learning, due to its great documentation and plethora of available tools. It supports many simulators for quantum algorithms testing and also includes the necessary functions to support quantum circuit execution to a variety of quantum hardware ranging from superconducting qubits QCs (IBM, Rigetti) to trapped-ion QCs (IonQ, AQT) etc. Xanadu developers have also created an experimental package, under the name Catalyst, that enables just-in-time (JIT) compilation of PennyLane programs \cite{ittah2024catalyst}. Furthermore, this package is equipped with advanced control flow routines that support both quantum and classical instructions, hence offering the opportunity for HPC-QC integration.  

\textbf{Cirq \cite{cirq_developers_2025_15191735}:} Cirq is a Python-based open-source library, designed by Google. It provides numerous abstractions for constructing and optimizing quantum circuits with respect to today’s NISQ era quantum hardware and simulators for accurate quantum algorithm validation. It comes with built-in functions which allow users to run their experiments on Google’s quantum processors and also with the necessary interface that enables users to execute their quantum circuits in other vendor’s (AQT, IonQ, Pasqal) quantum hardware. Furthermore, due to its complete compatibility with TensorFlow Quantum \cite{broughton2020tensorflow}, a library designed for building hybrid quantum-classical models with a focus on quantum data, it is feasible to integrate QCs with HPCs by exploiting Cirq. 

\textbf{Braket \cite{awsBraket}:} AWS Braket is a cloud-based service developed by Amazon. It is fully interconnected with the AWS service, thus helping users create quantum circuits, deploy and execute them in different types of quantum simulators and real hardware provided by IonQ, IQM and Rigetti. Furthermore, this dependency from AWS services provides Braket users the opportunity to develop hybrid quantum-classical algorithms and deploy them into hybrid HPC-QC systems without any struggle. The main reason for that is that AWS comprises a complete tool kit for HPC programming and resource management, like the ParallelCluster and the Parallel Computing Services, that integrate flawlessly to the AWS core. 

\textbf{Azure Quantum \cite{Microsoft_Azure_Quantum_Development}:} Azure Quantum is a cloud-based service developed by Microsoft. Azure Quantum is part of Microsoft’s Azure ecosystem and comprises of a wide spectrum of frameworks and platforms, like the Quantum compute platform. By exploiting these platforms users are able to construct and execute their quantum circuits to simulators and real hardware devices provided by Microsoft \cite{Microsoft_Azure_Quantum_Development}, by either using Python or Q \# as their language of preference. Q \# is a high-level, open-source programming language developed by Microsoft for writing quantum programs and it is included in the Microsoft’s Quantum Development Kit \cite{Microsoft_Azure_Quantum_Development}. Like Amazon Braket, Azure Quantum users have the opportunity to utilize the Azure platform and its services in order to create hybrid HPC-QC workloads and deploy them into state-of-the-art systems. 

\textbf{Rigetti QCS \cite{karalekas2020quantum}:} Rigetti QCS, is a Quantum Cloud-based service provided by Rigetti. It provides the necessary tools for users to construct and execute their quantum circuits on Rigetti’s quantum simulators and superconducting quantum hardware, via the Forest SDK \cite{smith2016practical}, by using either Quil or Python. Quil, which stands for Quantum Instruction Language, is a programming language for developing quantum programs, via the Quil SDK, and execute them in Rigetti’s QPUs. Additionally, Rigetti’s developers have created pyQuil, a python library that allows users to generate Quil programs from quantum gates and classical operations written in Python. Also, it comprises the necessary abstractions, like low-latency access, security and scalability of both classical and quantum resources, in order to offer users a reliable framework for HPC-QC hybrid programming.

Apart from the aforementioned QC frameworks that, nowadays, include the necessary routines for HPC-QC integration, some specifically-designed HPC-QC platforms have been developed also.  

\textbf{XACC \cite{mccaskey2020xacc}:} XACC, which stands for eXtreme-scale Accelerator programming framework, is a system-level software framework designed by A.J. McCaskey et. al. This framework is built upon a service-oriented architecture to assist users program, compile and run their hybrid workloads into HPC-QC systems. It is structured in a hardware-agnostic manner in order to provide support for both current and future HPC-QC integrated systems. XACC is implemented as a C++ framework, mainly due to C++ proven efficiency and performance, but is supports programming in other high-level languages, like Python, Julia, etc. through language binding libraries.  

\textbf{CUDA-Q \cite{the_cuda_q_development_team_2025_15407754}:} CUDA-Q is an open-source platform for quantum development, designed by NVIDIA. Alike the aforementioned QC frameworks, CUDA-Q is developed mostly to support hybrid classical-quantum large-scale workloads. The platform’s architecture allows computation on both classical devices, like CPUs and GPUs and on quantum hardware resources. Like XACC, it is constructed in a QPU-agnostic manner, hence integrating with all qubit technologies. Additionally, it offers a variety of tools for large-scale systems error-correction and also it supports GPU-accelerated simulations when the corresponding quantum resources are not available. 

\textbf{Pilot-Quantum \cite{mantha2024pilot}:} Pilot-Quantum is an open-source middleware for monitoring and controlling resources and workloads across HPC-QC systems, designed by P. Mantha et. al. Its implementation is based on the P* model \cite{luckow2012p}, which is a minimal, but complete model for Pilot abstractions \cite{luckow2012towards}, introduced by A. Luckow et. al. Pilot-Quantum comprises routines for effective resource and workload management along with scheduling in the application layer. Moreover, it was designed so that it could incorporate seamlessly with most available quantum programming frameworks, like Qiskit, Cirq, Pennylane etc.

%%%%%%%%%%%%%%%%%%%%%%%%%%%%%%%%%%%%%%%%%%%%%%%%%%%%%%%%%%%%%%%%%%%%%%%%%%%%%%%%%%%%%%%%%%%%%%%%%%%%%%%%%%%%%%%%%%%%%%%%%%%
%%%%%% 

\begin{figure*}[t!]
    \centering
    \includegraphics[width=0.80\textwidth]{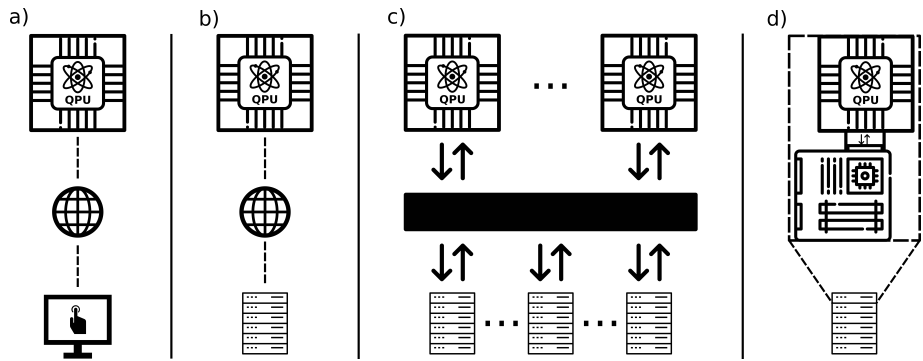}
    \caption{a) Standalone QPU - Loose integration: User interacts with a single QC through a web interface. b) Co-location - Loose Integration: The HPC interacts with a single QPU. They are two distinct infrastructures which are physically near and communicate through common local network or even apart from each other and communicate via cloud services c) Co-location - Tight Integration: The HPC system interacts with multiple QPUs. They are physically near but still separate hardware infrastructures which communicate via common hardware high-speed interconnects d) On-node - Tight Integration: The QPU is now embedded inside the classical computing node in the form of an external co-processor like a GPU.}
    \label{fig1}
\Description{Four schematic diagrams of HPC–quantum integration architectures. 
(a) Standalone QPU with loose integration, where the user accesses a quantum processor through a web interface. 
(b) Co-located loose integration, with an HPC system connected to a single QPU via a local or cloud network. 
(c) Co-located tight integration, where the HPC system connects to multiple QPUs using high-speed hardware interconnects. 
(d) On-node tight integration, where a QPU is embedded as an external co-processor within a classical computing node.}

\end{figure*}

%%%%%%%%%%%%%%%%%%%%%%%%%%%%%%%%%%%%%%%%%%%%%%%%%%%%%%%%%%%%%%%%%%%%%%%%%%%%%%%%%%%%%%%%%%%%%%%%%%%%%%%%%%%%%%%%%%%%%%

\section{Hardware Level Integration}
\label{sec:hardware_interface}
\subsection{Architectural Approaches}
\label{architectural}
Towards the creation of Hybrid HPC-QC systems, the approach that will be followed on the hardware-level integration of the corresponding components is crucial, as it will affect several aspects of the whole system, its total complexity and performance, and will also define the requirements and the needed components of the software stack for software-level interfacing.

The integration of HPC and QC systems can be clustered, among others, into three (3) main distinct categories, based on the followed microarchitecture and thus the relative location between the two physical subsystems and their interaction. Those three ($3$) categories are: 1) Standalone, 2) Co-located, and 3) On-node integration \cite{elsharkawy2023integration, rallis2025hardware}.

\textbf{1) Standalone:} This $1^{st}$ category involves a scenario where the QPU functions independently, without any automated connection to a HPC system, aligning with the principles of \textbf{loose integration model}. A traditional computing system is still required to enable user interaction with the QPU, typically through a web-based interface, as illustrated in Fig.~\ref{fig1}(a). In such setups, users are responsible for manually preparing their computational tasks and oversee the whole QC workflow. This includes formulating the problem, selecting or designing an appropriate algorithm \cite{rohe2024problem}, and translating the quantum algorithm into a high-level quantum-compatible representation, most often a quantum circuit, using high-level programming languages. The quantum circuit may then be optimized and transpiled into hardware-specific instructions for execution on a designated QPU, which returns the output for further processing \cite{javadi2024quantum}. This model lacks any automated HPC-QC hybrid workflow support and offers minimal system integration. Nevertheless, it is particularly well-suited for quantum algorithm development, testing, and evaluation, as well as for exploring the properties and performance characteristics of various QPUs and qubit technologies \cite{weder2020integrating}.

\textbf{2) Co-location:} The $2^{nd}$ category describes systems in which HPC and QC hardware are placed close to each other. Co-location typically implies that both resources reside in the same physical environment or communicate over a shared network infrastructure. In some cases, they may also be housed separately but interact through cloud-based connections. This category includes two ($2$) primary configurations: one where HPC and QC systems are physically co-located, and another where they are network-connected but not necessarily in the same physical location.

The first configuration, co-location via a shared communication network, adheres to the principles of the \textbf{loose integration model}. In this setup, the HPC system, utilizing its classical computing nodes, can interact with a single QPU, as illustrated in Fig.~\ref{fig1}(b), enabling the execution of hybrid quantum-classical workloads. The general workflow mirrors that of the standalone approach: it begins with preparing the quantum algorithm and ends with post-processing the results. However, a notable advancement in this configuration is the automation of several steps that previously required manual intervention \cite{humble2021quantum}. Additionally, this type of interface supports the execution of hybrid iterative algorithms which rely on frequent data exchanges between classical and quantum components. Despite these enhancements, achieving efficient automated communication demands a more advanced and intricate software stack to coordinate system resources effectively. Given the challenges associated with deploying and maintaining QPUs locally, cloud-based communication is increasingly common. Many QPUs are now provided through Quantum-as-a-Service (QaaS) models \cite{ahmad2024reference}, allowing remote access rather than local integration. However, this approach must contend with network-induced latency, which can hinder overall system performance and create bottlenecks. Moreover, if communication occurs over public or cloud-based networks, data security becomes a significant concern \cite{golec2024quantum}.

The second configuration involves the physical co-existence of resources and aligns with the principles of the \textbf{tight model of integration}. In this arrangement, multiple QPUs are physically situated alongside classical computing nodes, as depicted in Fig.~\ref{fig1}(c). These QPUs—potentially sourced from different vendors—are interconnected and communicate both with each other and with classical components through specialized, low-latency hardware interfaces. This physical proximity allows for the expansion of available quantum computational capacity and facilitates efficient data exchange, fostering deeper collaboration between classical and quantum processes while significantly minimizing latency. By incorporating a greater number of QPUs, this architecture supports more effective task distribution and the execution of complex quantum algorithms across multiple units, in a fashion akin to classical parallel computing \cite{pastor2024circuit}. However, due to its tightly integrated and heterogeneous nature, this setup demands a more advanced and robust software stack. The system must be capable of managing diverse QPU technologies, coordinating resource scheduling, and ensuring smooth, high-throughput data communication to support seamless operation.

\textbf{3) On-node Integration:} The $3^{rd}$ category refers to the direct integration of QPUs within HPC nodes themselves, as illustrated in Fig.~\ref{fig1}d), which is practically the definition of \textbf{tight integration}. This approach mirrors the integration of other hardware accelerators like GPUs or TPUs and follows the principles of tight coupling. It represents the most advanced vision for hybrid HPC-QC architectures, encompassing implementations that range from QPUs directly mounted on motherboards to more sophisticated and futuristic designs utilizing chiplets that scale-up QPUs and combine classical and quantum components into a unified system \cite{ruefenacht2022bringing, 10.1109/MICRO56248.2022.00078}. In this configuration, QPUs work in close conjunction with classical processors, even as low as at the instruction level, enabling seamless execution of hybrid algorithms where classical and quantum operations interact automatically. This effectively transforms QPUs into powerful, application-specific accelerators optimized for particular problem domains that can leverage their unique capabilities \cite{chen2024multi}. The physical proximity within this setup allows for real-time quantum-classical computation, significantly reducing latency issues that arise in other models where communication occurs over external networks. Moreover, it inherently enhances data security by eliminating the need for public or cloud-based data transmission. Despite its advantages, on-node integration is the most technically demanding model. It requires highly complex hardware and software frameworks to manage resource coordination and ensure smooth communication. The integration challenge is further intensified by the technological intricacies of qubit fabrication and their dependence on ultra-low temperature operation, especially when QPUs are to be packaged alongside classical components.

A summarized version of those four (4) different setups that lay in the three (3) described categories, can be viewed in Tab.~\ref{tab:hpc_qc_integration_comparison}.

\begin{table*}[htbp]
\centering
\caption{Comparison of HPC–Quantum Integration Architectures}
\label{tab:hpc_qc_integration_comparison}
%\renewcommand{\arraystretch}{1.1}
%\begin{adjustbox}{width=\textwidth}
{\small % Start smaller font size
\begin{tabular}{|p{2.65cm}|p{2.65cm}|p{2.65cm}|p{2.65cm}|p{2.65cm}|}
\hline
\textbf{Integration Type} & \textbf{Advantages} & \textbf{Limitations} & \textbf{Typical Interconnects} & \textbf{Use Cases} \\
\hline
\textbf{Standalone Integration} (Loose – Standalone) &
Simple setup. Easy access through cloud providers. Good for development and education. &
Very high latency. Manual workflows. No real-time hybrid execution. &
WAN / Internet, Web APIs. &
Quantum algorithm prototyping, educational purposes. \\
\hline
\textbf{Loose Co-Located Integration} (Loose – Co-located) &
Lower latency vs. Standalone. Local control of QPU. Basic hybrid workflows are possible. &
Network latency persists. Requires on-site QC hardware. Separate resource management. &
Ethernet, InfiniBand, Internet. &
Hybrid VQE/QAOA algorithms, on-premises hybrid experiments. \\
\hline
\textbf{Tight Co-Located Integration} (Tight – Co-located) & 
Low latency. Supports multi-QPU setups. Unified resource scheduling. &
Complex integration. Heterogeneous QPU vendor issues. Advanced orchestration is required. &
PCIe Gen 4/5, CXL, InfiniBand, Experimental Quantum Networking. &
Quantum chemistry, combinatorial optimization, multi-QPU hybrid HPC. \\
\hline
\textbf{On-Node Integration} (Tight – On-node) &
Near-zero latency. Real-time classical-quantum interaction. Enables adaptive hybrid workloads. &
Extreme hardware complexity. Cryogenic challenges. Still in research stage, no products yet. &
Direct PCIe, CXL, Cryo-CMOS controllers, Chiplet integration (future). &
Real-time hybrid HPC-QC, adaptive quantum error correction, next-gen HPC accelerators. \\
\hline

\end{tabular}
} % End smaller font size
%\end{adjustbox}
\end{table*}

\subsection{Physical interconnects between QPUs and HPC systems}

\textbf{Classical Peripheral Interfaces:} In many hybrid quantum-classical computing architectures, the quantum component, mainly the QPU and its control unit, is treated as a peripheral accelerator, much like a GPU. Following this approach, the connection between the HPC system and the QPU can be established using conventional high-speed interfaces such as PCI Express (PCIe). A notable example of this is the NVIDIA DGX Quantum system, which utilizes PCIe Gen 5 for QPU integration \cite{nvidiaDGXQuantum}. Beyond PCIe, another viable option for HPC-QPU interfacing is the Compute Express Link (CXL), which shares the same physical and electrical foundation as PCIe but offers enhanced capabilities. CXL supports high bandwidth and low latency communication while also enabling features like cache coherence and shared memory access, which are particularly beneficial for tightly coupled systems \cite{beck2024integrating}. These interfaces frequently include FPGA-based controllers that play a crucial role in generating the precise control pulses and readout signals needed for quantum operations \cite{stefanazzi2022qick, xu2023qubic}. FPGAs offer highly configurable, fine-grained control with accurate timing, allowing for reprogrammable and flexible coordination between classical and quantum units. However, to further improve performance, especially in terms of efficiency and integration density—Application-Specific Integrated Circuit (ASIC)-based quantum controllers are also being explored as an alternative to FPGAs \cite{nikbakhtnasrabadi2025quantum}.

\textbf{High-Speed Networks (InfiniBand, Ethernet):} As discussed earlier, most existing Hybrid HPC-QC systems currently adhere to the loose integration model. In such configurations, quantum computing components are often located remotely and typically housed within separate cryogenic infrastructure, and must communicate with HPC systems over a network. This interaction is commonly facilitated by high-speed networking technologies such as InfiniBand or advanced Ethernet connections. These networks handle the transmission of computational tasks and results between the HPC host and the QPU’s control unit \cite{beck2024integrating}. InfiniBand, in particular, is widely used due to its support for Remote Direct Memory Access (RDMA), offering exceptionally low latency and high throughput \cite{shpigelman2022nvidia}. This network-based approach is especially relevant in setups where quantum and classical components are not physically co-located but are instead interconnected through a local high-speed network.

\textbf{Cryogenic Wiring and Microwave Links:} Qubits can be fabricated using a variety of technologies. However, due to their extreme sensitivity to noise, most qubit implementations require ultra-low temperature environments, typically in the millikelvin range. For this reason, both the qubits and their associated control electronics are usually housed within cryostat. To accommodate this, various cryo-CMOS controller architectures have been proposed and studied, specifically designed to operate in cryogenic conditions near the qubits. These controllers are capable of generating control signals, amplifying signals when needed, especially during readout, and performing initial layers of local signal processing. A typical qubit readout employs a cryogenic low-noise amplifier, operating at very low temperatures, for signal boosting, apart from other isolating and filtering layers. Their proximity to the qubits helps streamline the physical interconnects, reduce latency and noise, enhance overall performance, and support the scalability of quantum systems \cite{staszewski2021cryo,oka2022cryo, hornibrook2015cryogenic, sebastiano2020cryo}.

To bridge the gap between room-temperature HPC electronics and cryogenically cooled qubits and their control units, two ($2$) primary types of physical interconnects are commonly employed: coaxial cables and waveguides. Coaxial cables are flexible and capable of transmitting both DC bias signals and high-frequency microwave pulses. In contrast, waveguides are rigid structures engineered specifically for the transmission of microwave signals at very high frequencies, offering advantages such as lower signal loss and reduced thermal conduction. These interconnects are essential for delivering microwave pulses, applying DC biases, and retrieving readout signals. Functionally, they form the physical layer connecting PCIe or FPGA-based controllers within the HPC system to the QPU. Typically, the control of a single qubit requires at least two dedicated input lines, along with additional input/output lines for readout operations \cite{das2024chip}. As quantum processors scale up, the number of required physical lines increases substantially, creating a significant bottleneck. To address this challenge, a variety of solutions have been proposed. These include the deployment of cryogenic RF switches and crossbar interconnect architectures \cite{potovcnik2021millikelvin, ciobanu2024optimal, li2018crossbar}, which aim to reduce the number of required connections. Additionally, frequency multiplexing is being explored to boost channel density—allowing multiple signals to share the same transmission medium. Promising developments include cryogenic electronic multiplexers \cite{acharya2023multiplexed} and photonic fiber links enhanced by the incorporation of wavelength division multiplexing (WDM) \cite{joshi2023scaling}, which, by using particular optical modulation schemes to encode various signals, may be able to combine numerous coaxial cables into a single optical fiber \cite{lecocq2021control}.

\textbf{Quantum Networking:} As discussed in Section~\ref{architectural}, hybrid quantum-classical systems can incorporate multiple QPUs, potentially from different vendors. To enable tighter integration and harness the combined computational power of these QPUs, especially for solving more complex problems that demand a larger number of qubits, direct communication between QPUs becomes essential. This form of inter-QPU communication, occurring without intervention by classical HPC components, is known as Quantum Networking. Quantum networking facilitates the transfer of quantum states between QPUs through entangled links, and the type of interconnection used depends on both the integration approach and the underlying qubit technology \cite{escofet2023interconnect}. Several promising technologies are under development to support this. For instance, photonic-based links using waveguides or optical fibers are suitable for transmitting quantum states across long distances \cite{marinelli2023dynamically}. In scenarios where chips are physically close, inter-chip interconnects are used to directly link qubits across different chips. These can take the form of coaxial cables, superconducting transmission lines, or capacitive couplings via resonators, depending on the specific qubit implementation \cite{escofet2023interconnect}. Quantum teleportation offers a more advanced method, using entangled photon pairs to transfer quantum information remotely between qubits without physical transmission of the particles themselves \cite{hu2023progress}. Another approach, ion shuttling, is used in ion-trap systems and involves the physical movement of ions within the cryogenic setup to bring qubits into close enough proximity for interaction \cite{schoenberger2024shuttling}. For the realization of quantum networks, the development of quantum transducers is critical. This technology allows for coherent conversion between different quantum information carriers, enabling cross-platform entanglement and integrating diverse QPUs into a unified quantum-classical computing environment \cite{balram2022piezoelectric}. These interconnection techniques are still in the experimental phase, but they lay the foundation for the next generation of quantum computing—where information will be exchanged through qubits rather than classical bits. This will enable distributed or parallel quantum computation and potentially lead to the formation of quantum local area networks (QLANs), where multiple entangled QPUs are orchestrated by HPC infrastructure \cite{azuma2023quantum, barral2024review}.

The use of those physical interconnects in HPC-QC integrated systems of different architectures, can be viewed in Tab.~\ref{tab:hpc_qc_integration_comparison}.

%%%%%%%%%%%%%%%%%%%%%%%%%%%%%%%%%%%%%%%%%%%%%%%%%%%%%%%%%%%%%%%%%%%%%%%

\section{Applications and Use Cases of HPC-QC interconnected systems}
\label{sec:applications}

While there is a rapid development of quantum computers technology, as highlighted in the Introduction section, we are currently in the so called NISQ era \cite{preskill2018quantum}. Despite the fact that current hardware reached a level of handling problems that require beyond $100$ qubits, their mapping yields deep circuits that due to noise lead to state decomposition and loss of information. On the other hand, classical HPC systems face significant challenges when tackling NP-hard and NP-complete problems, while it has been showcased that quantum computers can offer a significant advantage in solving such problems \cite{chatterjee2024solving}. As the size of these problems increases, the time required to solve them scales exponentially, demanding substantial computational resources. This rapid growth in execution time often renders solutions impractical for larger input dimensions. NP-hard and NP-complete problems are not just theoretical curiosities—they frequently arise in both research and industrial contexts, making their efficient resolution a critical concern across many domains \cite{callison2022hybrid}. To solve these kind of problems, hybrid quantum-classical approaches emerged \cite{callison2022hybrid, peruzzo2014variational} with the more prominent being the Variational Quantum Eigensolver (VQE) \cite{tilly2022variational,peruzzo2014variational} and the Quantum Approximate Optimization Algorithm (QAOA) \cite{farhi2014quantum}. The core idea behind these types of algorithms is to let quantum computers handle the tasks that are hard to solve on classical computers, and use classical ones for tasks like parameters adjustment, evaluation and pre/post processing, to name a few. In the following sections, we consider areas that HPC-QC interconnected systems present prominent applications. 
\subsection{Optimization}
\label{subsec:optimization}
Finding the global minimum or maximum of an objective function describing a problem, is a major issue of traditional optimization techniques like Bayesian optimization \cite{klein2017fast,movckus1975bayesian} and genetic algorithms \cite{gerges2018genetic}. Classical approaches frequently getting trapped in local minima, especially when the problem size increases. As stated previously, quantum computers with the emergnce of variational algorithms, QAOA and quantum annealing \cite{finnila1994quantum, kadowaki1998quantum}, promise a compelling potential in solving complex combinatorial optimization problems. Quantum algorithms can accelerate the optimization process in large and complex areas like scheduling \cite{choi2020quantum}, logistics \cite{awasthi2023quantum}, transportation \cite{padmasola2025solving} and finance \cite{buonaiuto2023best,zaman2024po} to name a few. Typically, quantum optimization protocols, utilize the hybrid quantum-classical interconnection. The problem is expressed as an Ising Hamiltonian \cite{lucas2014ising}, the quantum algorithm prepares and evolves the states that corresponds to a solution, while the classical part is related to the estimation of the expected value and the parameter update. 

\subsection{Machine Learning}
\label{subsec:ml}
Quantum Machine Learning (QML) integrates quantum computing with Machine Learning (ML) to enhance various aspects of ML workflows. QML applications are often grouped by the type of input data, classical or quantum, and by the learning paradigm, i.e., supervised \cite{schuld2018supervised,havlivcek2019supervised}, unsupervised \cite{lloyd2013quantum}, and reinforcement learning \cite{meyer2022survey,saggio2021experimental}. In the current NISQ era, workflows where classical data are encoded into quantum circuits, utilizing various feature encoding techniques, gain increasing attention. By increasing the datasets scale these models can be highly benefited by the HPC integration, where data are distributed across several nodes. This hybrid integration can be further extended from the sole purpose of data handling and can be incorporated to the models themselves. Several approaches have been proposed, where a mixture of classical and quantum components synergize to create a seamless workflow \cite{biamonte2017quantum, hofmann2008kernel,schetakis2022review, ghasemian2022hybrid, cong2019quantum, gircha2023hybrid,rebentrost2014quantum}. Starting from the Multi-Layer Perceptron (MLP) quantum circuits can be integrated into the learning and prediction process by replacing whole layers of the model, or even specific neurons \cite{beer2020training,liliopoulos2025hybrid}. Generative Adversarial Networks (GANs) \cite{huang2021experimental,zoufal2019quantum} utilize a generator and a discriminator to create synthetic data that resemble the original ones, in cases where there is a shortage of data. Both of the generator and discriminator components can be replaced by Parameterized Quantum Circuits (PQCs), individually, both of them and as it is the case of MLPs, some parts of them. Another straightforward application for the hybrid approach is the reservoir computing \cite{dudas2023quantum, kornjavca2024large,zhu2024practical} paradigm. The straightforward approach in this case is that the reservoir output layer to be enhanced by quantum components. Yet, the greater advantage comes by replacing the reservoir itself with a quantum circuit that follows the systems properties and connectivity and thus correspond to a natural reservoir architecture \cite{suzuki2022natural}. Generally, for many ML workflows and models the hybrid architecture not only by replacing/enhancing classical components with quantum circuits but also by taking advantage of the effectiveness of quantum computers in tasks evolving linear algebra \cite{harrow2009quantum}.  

\subsection{Many-Body dynamics}
\label{subsec:many-body}
Feynman originally envisioned general-purpose quantum computers as powerful simulators for natural systems that are inherently quantum. With the advent of algorithms VQE and Quantum Monte Carlo adapted for quantum computers \cite{huggins2022unbiasing,kanno2024quantum, zhang2022quantum}, hybrid HPC-QC workflows have emerged, leveraging the quantum devices' efficiency in solving many-body problems alongside the scalability of classical high-performance computing. These workflows are particularly promising in quantum chemistry, where classical methods are nearing their computational limits. Even in the current NISQ era, growing research demonstrates the potential value quantum computers offer for challenging tasks such as electronic structure determination and ground-state energy calculations \cite{cao2019quantum, mcardle2020quantum,bauer2020quantum, kandala2017hardware, google2020hartree, motta2023quantum, aspuru2005simulated}. Robledo-Moreno et al. recently showcased that QCs, supported by large-scale classical resources, can achieve approximate but useful results in quantum chemistry problems that are beyond the reach of classical exact methods or standalone quantum processors \cite{robledo2024chemistry}. Drug discovery is another area where classical computing limitations appear. Again, the HPC-QC integration can offer advantages in crucial steps of the workflow such as the protein folding problem \cite{robert2021resource} and molecular docking \cite{papalitsas2025quantum}. Thus, the goal of such an HPC-QC framework is to enable the integration of quantum algorithms into simulations, where a particular part of the simulation can be offloaded to the quantum computer.

%%%%%%%%%%%%%%%%%%%%%%%%%%%%%%%%%%%%%%%%%%%%%%%%%%%%%%%%%%%%%%%%%%%%%%%%%%%%%%%%%%%%%%%%%%%%%%%%%%%%%%%%%%%%%%%%%%%%%%%

\section{Challenges and Limitations}
\label{sec:challenges}

Even though highly promising and under intensive investigation, the efficient and highly performant integration and operation of HPC-QC systems is affected by a significant and diverse challenges that need to be overcome. Those challenges are related with all the layers of HPC-QC system integration (application layer, middleware layer, hardware layer) and include among others, hardware maturity, data communication restrictions, software development, resource management, security, as well as cost and accessibility issues.

\subsection{Quantum Hardware Maturity} 
\label{subsec:maturity}
Quantum hardware is still continuously progressing and several different technologies are being tested and evaluated, targeting on providing more and of higher quality qubits. However, the existing quantum hardware technology in its whole, is still in the NISQ era, is accompanied with considerable limitations.

A distinctive challenges that is present in current hardware technology is qubit quality. For the time being, the provided qubits are extremely sensitive to external pertrubations from the environment, which causes their quantum states to decay overtime. This is the well-known problem of decoherence. The existing quantum devices of the NISQ era suffer from low coherence times, reducing the available duration of the quantum computation to the order of a few hundreds of microseconds and thus limiting the depth of the quantum circuits that can be executed \cite{delgado2025defining}.

Additionally, a significant source of noise, and thus errors is the application of quantum gates. Every quantum operation (gate) executed by a qubit introduces a small amount of noise which is accumulated and propagated through the quantum circuits. This is mainly attributed to the existing gate fidelities, which can ultimately lead to a cascade of errors \cite{heinz2021crosstalk}. The readout process and circuitry can also be responsible for the introduction of errors. Even if the quantum computation itself was performed within the required range of accuracy, the measurement process itself can lead to a wrong output \cite{karuppasamy2025comprehensive}.

As the qubit density increases, and qubits are naturally getting closer with one another, the crosstalk phenomenon becomes evident. This phenomenon introduces correlated errors, leading to corrupted quantum states that significantly affect the integrity of the quantum computation \cite{heinz2021crosstalk}.

Additionally, the scalability of the QPUs which mainly refers to the increase of the qubit count in a single quantum chip, is a very common and significant problem related to quantum hardware technology. At the current technology level, QPUs may include up to a few tens of hundreds of physical qubits, a number which is slowly but constantly increasing. The increase of this number is challenging across all qubit technologies, and with such small scale of QPUs, the current capabilities of solving large scale problems are restricted \cite{safi2023influence}. A significant limiting factor on the scalability of QPUs is the wiring complexity. Connecting control and readout lines from room-temperature classical electronics to each qubit inside a cryostat presents a significant wiring challenge that scales unfavorably with the number of qubits \cite{joshi2023scaling}.

Another significant problem related to quantum hardware is the connectivity. The intercommunication of qubits with one another is many times restricted, as each qubit can only interact with a small part of its neighbours. This can be problematic for quantum algorithms where the physical correlation is required between qubits that are distant. The use of SWAP gates, can provide a solution to this problem by bringing qubit states physically closer, however they add significant overhead which ultimately degrades the overall performance. For this reason, optimization is investigated towards the development of compilation techniques that minimize SWAP overhead \cite{zhu2025s}.

\subsection{Data Management and Communication} 
\label{subsec:managementCommunication}
Apart from the quality and the scalability of the hardware itself, the communication between the QPU and HPC parts, plays a significant role on the overall performance and thus the significance and usability of hybrid systems, and has its own challenges and bottlenecks.

As mentioned above, in Section \ref{sec:hardware_interface}, the latency and bandwidth can be significant problems in the communication between the HPC and QC systems. It is a crucial topic, taking under consideration the large volumes of data that needs to be exchanged between the two components, as well as the pre- and post- processing that is demanded, every time data is transferred from one subsystem to the other.
A significant amount of latency especially in hybrid algorithms that contain feedback loops, can possibly negate all the speedup gains provided by the QC. In such cases, the data exchange between a CPU and a QPU can become an intensive and time-consuming task, and its manipulation with efficient I/O strategies and techniques targeting to the minimization of overhead is a challenging but crucial step towards high-performance HPC-QC hybrid systems \cite{lubinski2022advancing}.

\subsection{Ecosystem Maturity} 
From a software point of view, there are still significant problems to be solved and challenges to be addressed, as the ecosystem for HPC-QC hybrid systems is far less mature compared to that of classical HPC systems.

At the current state, and in contrast with the existing HPC solutions, the software stack related to quantum computers and quantum programming is far less developed and established. This is reflected on the exploited programming tools, as quantum programming often utilizes programming languages that target ease of use and flexibility, such as Python or Julia, incorporating libraries and frameworks specific for Quantum Computing, while the majority of HPC applications employ programming languages like C and C++, or programming models and frameworks such as OpenMP and MPI, which target mainly on performance. Bridging this gap requires significant effort in developing robust, performance-oriented quantum software tools \cite{meller2025programming}. A significant problem that also affects the QC software stack and toolchain is the absence of standardization, which spans from as low as the hardware communication protocols, up to the software APIs and intermediate representations. This creates a fragmented ecosystem where each user provides its own heuristic solution, with no universal compatibility across multiple systems of different technologies, adding increased overhead and significantly reducing the re-usability of various software components. Some initial steps towards standardization have already been attempted through the realization of Quantum intermediate representations which are technology agnostic \cite{cardama2025review}.
At the higher level of the toolchain, another significant problem is the lack of a standardized and automatic methodology for mapping and translating classical problems into quantum formats. Breaking down a problem into parts, identifying which of those parts can benefit from quantum processing and how to map them onto quantum algorithms, as well as selecting which is the best quantum algorithm for a specific case, remains a highly challenging and active area of research \cite{volpe2024predictive}.

\subsection{Resource Management and Scheduling} 
In the area of middleware, the resource management and job scheduling in hybrid systems that contain not only heterogeneous computational resources, but go even beyond conventional by including also multiple QPUs from different vendors, are complex and challenging tasks.

In their current form, QPUs provided by vendors are being accessed through web APIs with proprietary queuing systems, which makes it really challenging to be integrated with the existing sophisticated schedulers used by the conventional HPC systems, such as SLURM. Thus, the need for development of specific interfaces or plugins for existing HPC schedulers that will allow the seamless integration and coordination of diverse QPU resources with classical resources is evident. In such efforts, another significant problem that needs to be taken under consideration and requires specific handling, is the possible workload imbalance which is introduced with the use of QPUs. Different QPU technologies can have very different operation times, which makes it hard for traditional HPC schedulers to operate effectively, as they often assume more predictable job durations. On the other hand, simple co-scheduling strategies, like the exclusive allocation of a QPU to an HPC job for its whole duration, can be proven extremely inefficient by leading to underutilization of either the classical or the quantum node \cite{viviani2025assessing, giortamis2025qos}. Thus, the development of scheduling policies which are able to cover dependencies between heterogeneous HPC and QC resources and also target optimization for overall performance and ensure fairness among users is an active and challenging area of research

\subsection{Cost and Accessibility} 
Finally, among others, a major challenge that is simultaneously a significant obstacle in the broad development and adaptation of QCs, and consequently to actual realization of high-performance hybrid HPC-QC systems, is the increased financial investment required for both owning and accessing them.

Both HPC and QC components are made of high-cost technologies. Especially QC components are inherently expensive technologies, as they are relatively novel, require complex fabrication processes, expensive specialized hardware and demand environmental controls in order to operate, such as cryogenics for a significant amount of QPU types. All this, increases the cost of operation and also the cost of maintenance. For this reason cloud computing is currently dominating the area of QCs and also HPC-QC systems, where access is provided through cloud platforms by companies, significantly lowering access costs and making those technologies available to a broad audience \cite{golec2024quantum}.

However, apart from the use of physical QPUs in QC or hybrid HPC-QC systems, the simulation of such workflows and quantum tasks can also be considered expensive. Simulating quantum computations on classical computing systems can be extremely resource intensive even for small scale problems, a fact that significantly slows the development and investigation of new quantum algorithms \cite{zhou2020limits}. The area of simulation of quantum operations with classical computing systems is a challenging area of research where also application specific hardware that emulates quantum operations is explored and tested \cite{fyrigos2023quantum}.

\section{Discussion and Conclusions} 
\label{sec:conclusions}
The integration of High-Performance Computing (HPC) with Quantum Computing (QC) represents a transformative shift in computational systems, combining the strengths of classical processing power with quantum-related advantages. Throughout this manuscript, several aspects of HPC-QC interfacing have been analyzed, including system architecture models, the evolving software and middleware ecosystem, hardware integration strategies, and representative application domains. These hybrid systems combine and leverage the complementary strengths of HPC and QC, establishing a framework for solving problems that neither modality could solve efficiently on its own.

Architectural approaches on HPC-QC integration span from loosely to tightly coupled designs. At one end, the loosely-coupled configurations treat QPUs as external systems, accelerators accessed through high-speed networks or cloud services, offering simplicity in implementation, at the cost of increased communication latency. On the other hand, co-located models bring quantum and classical resources in closer physical proximity, to reduce latency and improve throughput. At the extreme case, on-node integration envisions the incorporation of a QPU directly into an HPC node, enabling near-zero latency and real-time quantum-classical interactions. Each model entails its own distinct trade-offs in terms of latency and complexity: While loosely integrated systems are currently the most practical, the on-node integration paradigm is viewed as the long-term ideal in terms of performance, despite the existing engineering challenges.

In terms of software, the ecosystem that supports hybrid HPC-QC systems has matured significantly in recent years. High-level programming frameworks are constantly evolving and have already expanded the toolkit for developing hybrid algorithms, providing to the developers easily accessible, tunable, and standardized interfaces. Simultaneously, several software platforms, such as NVIDIA's CUDA Quantum and XACC frameworks have emerged, working towards the effective unification of classical and quantum computing workflows, allowing quantum accelerators to be included within traditional HPC applications. Adding to that, at the lower, middleware layer, dedicated solutions have already appeared and are constantly developing, coping with the significant challenges of resource management orchestration, job scheduling, and task execution across both HPC and QC resources. Those middleware tools showcase the significance of solid combined workload management in achieving a seamless integration. Together, the advances in frameworks and middleware are steering the community toward a more unified software stack for hybrid HPC-QC computing, although further improvements in interoperability, developer tools, and programming abstractions are still needed to reach the level of maturity seen in established classical HPC software.

Nonetheless, significant technical and operational challenges remain before HPC-QC systems can reach their full potential. Existing quantum computing hardware is still at the NISQ era, characterized by limited qubit counts, short coherence times, high error rates, and restricted qubit connectivity. These limitations significantly reduce the size and complexity of quantum subroutines that can be reliably executed on current hardware, and showcase the need for more efficient error correction algorithms and less noise-prone future quantum devices. The software ecosystem is also still at an early stage of development, far behind the maturity level of that of conventional HPC systems. Even though constantly evolving, it still lacks universal standards and full interoperability between platforms. 
From an operational point of view, integrating quantum processors into HPC infrastructures also presents non-trivial challenges. Issues such as the need for cryogenic environments and specialized control electronics for many QPU technologies, the complexity of co-scheduling quantum jobs alongside classical tasks, and the increased cost and expertise required to maintain quantum hardware, all pose practical obstacles. These challenges emphasize the necessity for continued research, engineering innovation, and investment in both hardware and software to overcome current bottlenecks and ensure that hybrid HPC-QC systems can be scaled up and used reliably.

Looking forward, ongoing and future developments promise to further tighten the integration between HPC and quantum platforms. On the hardware side and mainly on the part of peripheral circuits and systems that are responsible for connecting and controlling QC hardware, new technologies are constantly being explored to enable tighter interconnection with lower latency in hybrid systems. In addition, quantum networking technologies are expected to play a significant role in the longer term by allowing multiple QPUs to be linked into a unified computational fabric, extending the reach of hybrid architectures beyond a single QPU. Simultaneously, the software and middleware layers are about to evolve, following the well-established paradigms of conventional supercomputing. In addition, more sophisticated job schedulers and workload managers tailored for quantum-enhanced HPC clusters, improvements in real-time execution of hybrid quantum-classical workflows, and standardized application programming interfaces (APIs) are expected to emerge in the near future. These developments will improve the portability, usability, and efficiency of HPC-QC systems, making them more accessible. Moreover, as quantum device fabrication and gate fidelity improve, leading to lower hardware error rates, QC technology is expected to migrate from the NISQ, to the FTQC (Fault-Tolerant Quantum Computing) era in the forthcoming years. This change toward fault-tolerant QPUs, can ultimately lead to large-scale quantum computations, enabling far deeper quantum circuits and more complex algorithms to be executed reliably as part of HPC workflows.

Hybrid HPC-QC systems are set to become necessary tools for a range of high-impact applications as they evolve. Practically, combining HPC with quantum accelerators holds great promise for solving large-scale optimization problems (e.g. in logistics or portfolio optimization) and enhancing machine learning tasks, as well as for enabling more accurate simulations of many-body quantum systems in chemistry and materials science that exceed the practical bounds of conventional computing. Early demonstrations in these domains have already shown that quantum co-processors, when used together with classical HPC resources, can possibly provide computational speedups or increased result accuracy for certain problem categories. As hybrid architectures mature, we can expect their impact to broaden across scientific research and industry, offering new capabilities for tackling challenges that were previously considered intractable. Ultimately, the bridging of high-performance classical computing with increasingly powerful quantum computing is likely to redefine the boundaries of computational capability. By harnessing the unique strengths of both paradigms in a unified system, future HPC-QC platforms promise to push far beyond today’s computational frontiers.

%%%%%%%%%%%%%%%%%%%%%%%%%%%%%%%%%%%%%%%%%%%%%%%%%%%%%%%%%%%%%%%%%%%%%%%%%%%%%%%%

\begin{acks}
This research has been supported by the project “A catalyst for EuropeaN ClOUd Services in the era of data spaces, high-performance and edge computing (NOUS)”, Grant Agreement Number 101135927. Funded by the European Union's HORIZON-CL4-2023-DATA-01 call, views and opinions expressed are, however, those of the authors only and do not necessarily reflect those of the European Union. Neither the European Union nor the granting authority can be held responsible for them.
\end{acks}

\bibliographystyle{unsrtnat}
\bibliography{sample-manuscript}

%\appendix

\end{document}